\documentclass[10pt,aas2pp4]{aastex}

\usepackage{psfig}

\def\etal{{et\,al.}}
\def\msun{M$_{\odot}$}

\def\degs{\ifmmode ^{\circ}\else$^{\circ}$\fi}
\def\amin{\ifmmode ^{\prime}\else$^{\prime}$\fi}
\def\asec{\ifmmode ^{\prime\prime}\else$^{\prime\prime}$\fi}
\newbox\grsign \setbox\grsign=\hbox{$>$}
\newdimen\grdimen \grdimen=\ht\grsign
\newbox\laxbox \newbox\gaxbox
\setbox\gaxbox=\hbox{\raise.5ex\hbox{$>$}\llap
     {\lower.5ex\hbox{$\sim$}}}\ht1=\grdimen\dp1=0pt
\setbox\laxbox=\hbox{\raise.5ex\hbox{$<$}\llap
     {\lower.5ex\hbox{$\sim$}}}\ht2=\grdimen\dp2=0pt

\def\grs{GRS 1915+105}

\def\1e17{1E 1740.7-2942}
\def\gr17{GRS 1758-258}
\def\4u{4U~1630-47}

\begin{document}


\title{The 590\,days long term periodicity of the microquasar GRS 1915+105}

\author{A. Rau\altaffilmark{1,2}, J. Greiner\altaffilmark{1,2}, M. L. McCollough\altaffilmark{3}}

\altaffiltext{1}{Astrophysical Institute Potsdam, An der Sternwarte 16, 14482 Potsdam, Germany}
\altaffiltext{2}{Max-Planck-Institute for extraterrestrial Physics, Giessenbachstrasse,
  85748 Garching, Germany; arau@mpe.mpg.de, jcg@mpe.mpg.de}
\altaffiltext{3}{National Space Science and Technology Center, 320 Sparkman
  Drive, Huntsville, AL 35805; mccolml@head-cfa.harvard.edu}

 
\begin{abstract}
We report on the discovery of a 590\,days long term periodicity in the
hard X-ray component of the microquasar \grs\ found from a
comprehensive study of more than four years of {\it RXTE}
observations. The periodicity is also observed in the hard X-ray flux
observed by BATSE, and in the radio flux as seen with the {\it Green
Bank Interferometer} and the {\it Ryle Telescope}. We discuss various
possible explanations, including the precession of a radiation induced
warped accretion disk.
\end{abstract}


\section{Introduction}

The galactic X-ray binary system \grs\ is the most prominent
microquasar showing dramatic variability in its light curve (Greiner
\etal\ 1996), quasi-periodic oscillations, phase lags and coherence
behavior (Morgan \etal\ 1997, Muno \etal\ 2001). Microquasars are
thought to be down-scaled analogs to quasars harboring a stellar mass
black hole which accretes matter from a companion star. They show
superluminal ejections (Mirabel \& Rodriguez 1994) making the sources
potential laboratories for studying accretion and relativistic jets
near black holes (Mirabel \etal\ 1992). \grs\ is the most energetic
object known in our galaxy with a luminosity of $\sim5\cdot 10^{39}$
erg/s in the high state and $\sim 10^{38}$ erg/s in the low state. The
binary system, located at a distance of $\sim$12~kpc, contains a
1\,\msun\ late type giant (Greiner \etal\ 2001a) in a 33.5\,days orbit
around a 14\,\msun\ black hole (Greiner \etal\ 2001b), making it the
most massive stellar black hole known.

Originally discovered by GRANAT (Castro-Tirado \etal\ 1992) \grs\ has
been extensively monitored by the {\it RXTE} satellite since 1996, and
a number of investigations of these data have been published
(e.g. Greiner \etal 1996, Morgan \etal\ 1997, Belloni \etal\ 1997 \&
2000, Muno \etal\ 1999 \& 2001, Rau \& Greiner 2003; hereafter RG03).
 
In a recent paper (RG03) we presented a comprehensive study of the
X-ray spectral behavior of \grs\ in the $\chi$-state (Belloni \etal\
2000) for over nearly four years of observation with {\it
RXTE}. Within this study, a remarkable long term periodicity was found
which was mentioned briefly by Kuulkers \etal\ (1997). They found a
19\,month long term variability comparing the X-ray outbursts observed
with BASTE and {\it RXTE} data from August 1992, March 1994, October
1995 and May 1997. This is consistent with our presented result.

A number of X-ray binaries, including LMXB as well as HMXB, exhibit
long term variability in their X-ray flux with periods remarkably
longer then their orbital period. Tananbaum \etal\ (1972) found a
35\,days period in Her X-1 which they explained by the precession of a
warped accretion disk in the system. The same process is used to
describe the long term variations in SMC X-1 (Wojdowski \etal\ 1998),
LMC X-3 (Cowley \etal\ 1991) and LMC X-4 (Heemskerk \& van Paradijs
1989).

In this letter we discuss the detection of a new long term periodicity
in \grs.

\section{Data reduction and analysis}

The analysis presented here is based on the data reduction described
in full detail in RG03. We investigated public RXTE data from \grs\
from November 1996 to September 2000 obtained from the HEASARC.  We
analyzed data of the PCU0 detector of the Proportional Counter Array
(PCA) from 3--25\,keV and data of the High Energy X-ray Timing
Experiment (HEXTE) cluster 0 from 20--190\,keV for 139 $\chi$-state
observations from 89 different days. $\chi$-states correspond to the
low/hard state of \grs, exhibiting relatively low flux from the
accretion disk and showing a hard power law tail. These states are
characterized by the lack of obvious variations in light curve and
spectrum and are connected by continuous radio emission of varying
strength.

For the spectral fitting in XSPEC 11.0 (Arnaud 1996) we used a model
consisting of (i) photo-electric absorption (WABS), (ii) a spectrum
from an accretion disk consisting of multiple blackbody components
(DISKBB) and (iii) a power law spectrum reflected from an ionized
relativistic accretion disk (REFSCH; Fabian \etal\ 1989, Magdziarz \&
Zdziarski 1995).

\begin{figure}
\begin{center}
\psfig{figure=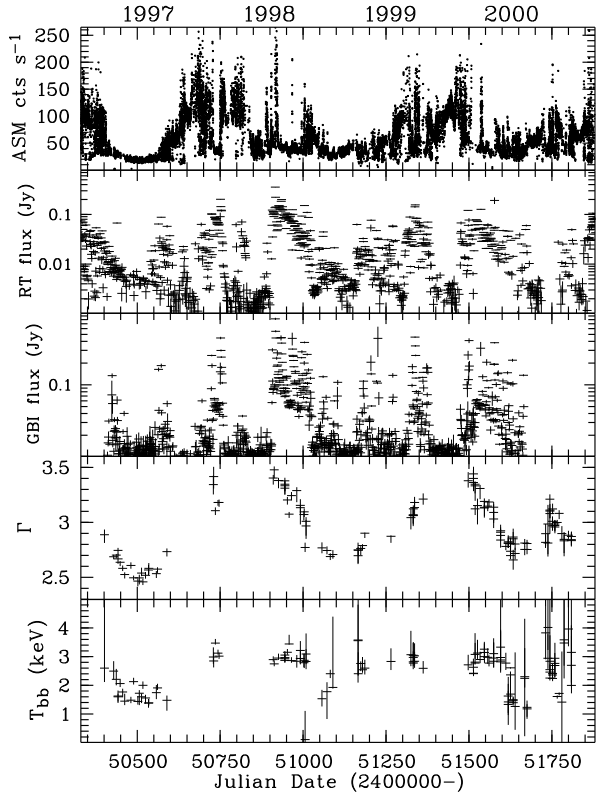,width=8.0cm,%
          bbllx=1.9cm,bblly=3.5cm,bburx=19.0cm,bbury=26.1cm,clip=}\par
 \caption[]{\begin{it}1.5--12\,keV {\it RXTE} All Sky Monitor (ASM)
          count rate (top panel), 15\,GHz {\it RT} (2nd from top) and
          2.25\,GHz {\it GBI} flux (3rd) from November 1996 to
          September 2000. ({\it GBI} was shutdown some time before
          JD $\sim$2450400 and following JD $\sim$2451660.) The lower
          panels show the power law slope, $\Gamma$, (4rd) and the
          accretion disk temperature, $T_{bb}$, (5th) for all observed
          $\chi$-states of \grs. Error bars are 1$\sigma$ for each
          parameter of interest.\end{it}
\label{fitres}}
\vspace{-1.0cm}
\end{center}
\end{figure}

The analysis of the X-ray model parameters for all $\chi$-state
observations of \grs\ reveals a remarkably periodic behavior of the
power law slope, $\Gamma$, which dominates the hard X-ray spectrum. It
shows a nearly sinusoidal variation (Fig.~\ref{fitres}, see also RG03)
on a time scale much longer then the orbital period, $P_{orb}$. The
recently found correlation of the power law slope and the radio flux,
$F_R$, in the $\chi$-states (RG03) suggests that one should also
search for a long term periodicity in the radio data of the {\it Green
Bank Interferometer} ({\it GBI}; 2.25 and 8.3\,GHz) and the {\it Ryle
Telescope} ({\it RT}; 15\,GHz).

\begin{figure}[h]
 \begin{center}
 \vbox{\psfig{figure=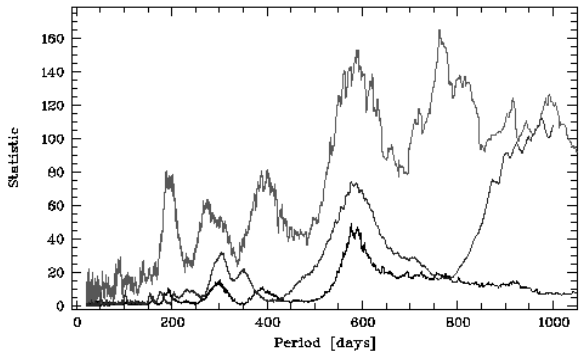,width=8cm,%
          bbllx=7.8cm,bblly=2.5cm,bburx=18.3cm,bbury=20.0cm,angle=-90,clip=}}\par
\caption[Varianzanalyse der Potenzgesetzkomponente]{\begin{it}Analysis
of variance of the 8.3\,GHz GBI radio flux (top curve), the
20-200\,keV BATSE X-ray flux (middle) and the power law slope (bottom)
for JD 2450300-2451800. The y-axis is in units of the test statistic
of the Fisher-Snedecor distribution function. The test statistic has
larger values for the radio flux due to the higher number of {\it GBI}
observations compared to BATSE and {\it RXTE}. A maximum at
590$\pm$40\,days is clearly seen in each of the three
graphs. \end{it}}
\label{tsa_alpha}
\vspace{-0.7cm}
\end{center}
\end{figure}

\begin{figure}[h] \begin{center}
 \vbox{\psfig{figure=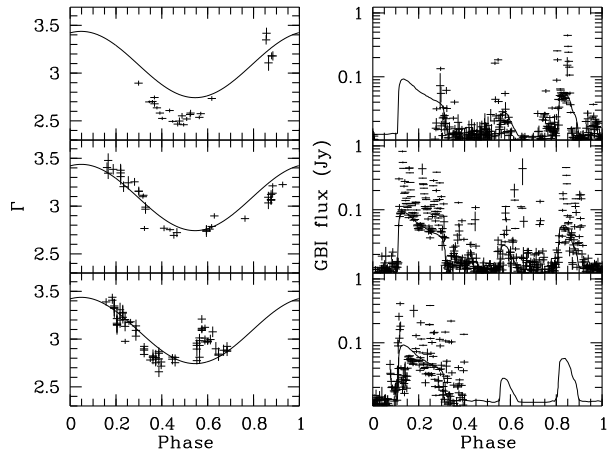,width=8cm,%
 bbllx=1.1cm,bblly=10.5cm,bburx=19.5cm,bbury=24.5cm,clip=}}\par
 \caption[Phasengefaltete Potenzgesetzsteigung und 2.25\,GHz
 Fl\"usse]{\begin{it}Power law slope, $\Gamma$, of all analyzed
 $\chi$-state observations (left) and of the 2.25\,GHz radio flux from
 {\it GBI} (right) from JD 2450350--2450940, JD 2450940--2451530 and
 JD 2451530--2452120 (top to bottom panels). Each panel covers a
 590\,days period and JD=2450225 was chosen as the zero point for
 plotting reasons only. The solid lines visualize the best fitting
 sinusoidal function (left) and illustrate the repeating outbursts in
 the radio flux (right). \end{it}}
\vspace{-1.0cm}
\label{phase}
\end{center}
\end{figure}

For the determination of a cycle duration an analysis of variance for
several model parameters and X-ray and radio fluxes was performed
using the Fisher-Snedecor distribution function (Schwarzenberg-Czerny
1989). The results for $\Gamma$, the 20--200\,keV X-ray {\it
CGRO}/BATSE flux and the 8.3\,GHz {\it RT} flux are given in the
periodogram in Fig.~\ref{tsa_alpha}. It shows a maximum at a period,
$P_{long}$, of 590$\pm$40(FWHM)\,days.  $\Gamma$ and $F_R$ show
several identical local maxima around 200, 300 and 400 days which are
caused by the three major radio outbursts (see right panel of
Fig.~\ref{phase}) during a 590\,days period. This local maxima result
from the time separation of the outbursts at phase $\sim$0.6 and
$\sim$0.85 (200\,days), $\sim$0.2 and $\sim$0.6 (300\,days) and
$\sim$0.2 and $\sim$0.85 (400\,days). These local maxima in the
analysis of the variance of Gamma arise from a combination of two
factors.  First the chi-states are usually connected with radio
emission and second the chi-states have non-stochastic distribution
within the RXTE observations.  The later factor results in an uneven
coverage of data points within the time period.

   
Fig.~\ref{phase} shows the power law slope and the {\it GBI}
radio flux at 2.25\,GHz displayed in segments of 590\,days each. These
light curves demonstrate, that 590\,days is the preferred period rather
than 200,300 or 400 days. Even though the coverage with {\it GBI} data
is shorter than with {\it RXTE} observations the periodicity is
clearly visible in the radio outbursts. Both the broad outburst with
exponential cutoff at phase $\sim$0.2 and the shorter and steeper
outbursts repeat periodically.


The soft X-ray component did not show any periodicity in the
$\chi$-states. This is true for the ASM count rate and hardness ratio
as well as for the fit parameters of the accretion disk,
i.e. effective temperature and the inner disk radius.


\begin{figure}[h]
 \begin{center}
 \vbox{\psfig{figure=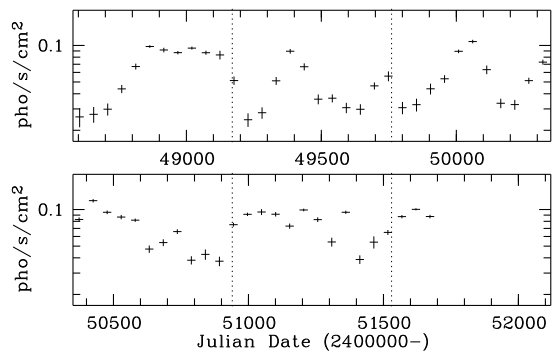,width=8cm,%
          bbllx=1.1cm,bblly=14.5cm,bburx=19.5cm,bbury=27.0cm,clip=}}\par
 \caption[]{\begin{it} BATSE 20-200\,keV flux from JD 2448580-2452120
 with a time binning of 57\,days. The dotted lines devide the light
 curve into 590\,day segments. \end{it}}
\vspace{-1.0cm}
\label{batse}
\end{center}
\end{figure}


The 590\, days period in the hard X-rays is also seen in the
20--200\,keV BATSE earth occultation fluxes
(Tab.\ref{tsaresultate}). However, the long term periodicity seems to
be variable on larger time scale. Fig.~\ref{batse} shows the
20-200\,keV light curve with a 52\,day (orbital precession period of
{\it CGRO}) binning. This binning ensures the removal of any
periodicity arising from the precession of the satellite
orbit. Although the 590\,day period is also visible when analyzing the
data from JD 2448300 to JD 2450300 (before {\it RXTE} observations
where performed, Tab.~\ref{tsaresultate}) the structures in the light
curve show remarkable differences and a somewhat shorter period of
545$\pm$25~days is found when analyzing all available BATSE data from
JD 2448300 to JD 2451800 together. A first hint for this variation is
seen when fitting the power law slope behavior with a simple
sinusoidal function with a 590\,days period (solid line in left panel
of Fig.~\ref{phase}). The minimum in he power law slope of the long
duration $\chi$-state from JD 2450400--2450600 lies outside the
minimum of the sinusoidal function (Fig.~\ref{phase} top). Also, an
additional maximum is seen around phase=0.6, particularly in the
bottom panel. The inpersistent behavior is not surprising, and
probably reflects the known transient behavior of \grs.


\begin{table}[h]
\caption[Ergebnisse der Varianzanalyse]{\begin{it} Results of the
analysis of variance for several model parameters and energy
ranges for the analysis over the time span of JD 2450300--2451800 with
exceptions explicitly stated otherwise. (n.d. = periodicity not detected)\end{it}}
\begin{center}
\begin{tabular}{lc}
\hline \hline
data & days \\
\hline \hline
$T_{bb}$ & n.d. \\
$\Gamma$ & 590$\pm$40 \\
$R$ & 590$\pm$40 \\
2.25\,GHz GBI flux & 600$\pm$40 \\
8.3\,GHz GBI flux & 590$\pm$40 \\
15\,GHz RT flux & 600$\pm$40 \\
GBI spectral index & 590$\pm$40 \\
1.5--12\,keV ASM cts/s & n.d. \\
20--200\,keV BATSE flux & 600$\pm$40 \\
20--200\,keV BATSE flux ({\tiny JD 2448300--2450300}) & 590$\pm$40 \\
20--200\,keV BATSE flux ({\tiny JD 2448300--2451800}) & 545$\pm$25 \\
\hline \hline
\end{tabular}
\label{tsaresultate}
\end{center}
\end{table}


\section{Interpretation}

The observation of the 590\,days periodicity in \grs\ suggests a
precessing warped accretion disk, analogous to what is seen in other
X-ray binaries. However, in all previous sources it was the soft X-ray
flux which showed the long-term modulation. In contrast, in the case
of \grs, it is the spectral slope of the power law and the radio
flux. The radio flux is connected with the hard X-ray component and
probably originates near the inner part of the disk. It is unlikely
that the periodic behavior of $F_R$ and $\Gamma$ can be explained by a
geometric effect of obscuration of the inner disk by the outer
accretion disk.

It has long been known (Katz 1973), that a precession of a warped disk
can be caused by tidal forces of the donor star. Assuming the
accretion disk consists of small concentric rings which are tilted
away from the orbital plane. Because of their fast rotation the rings
behave gyroscopically. Without interaction of the various individual
ring elements a retrograde precession around an axis perpendicular to
the orbital plane will be excited by the tidal torque. The precession
rate depends on the radius of the ring element which will lead to a tilted
disk. In a fluid disk, an internal torque between the ring elements
acts against the tilting. The resulting torque can enforce a
precession of the disk.  But the internal torque arises only if the
disk is inclined with respect to the orbital plain and is connected
with a dissipation of energy and a backward aligned movement of the
disk into the orbital plane.

For tidally forced precession, a correlation of $P_{long}/P_{orb}$ and
$q$ is theoretically expected but not found (Wijers \& Pringle 1999).
Neither the existence of a long term period nor the ratio
$P_{long}/P_{orb}$ are correlated with the mass ratios of the
components of the binary systems, ruling out tidal forces as the
generally dominant mechanism for the long term periodicities in these
systems. Thus, another mechanism is required which provides continuous
warping of the accretion disk. Radiation driven warping is one
possible mechanism (Petterson 1977, Pringle 1996, Maloney \etal\ 1996,
Ogilvie \& Dubus 2001). The outer, tilted part of the accretion disk
is irradiated by the central source. If the radiation is absorbed and
re-emitted parallel to the local disk gradient a torque due to gas
pressure affects the disk. Therefore, a small warp grows exponentially
if the luminosity is sufficient to depress the dissipative processes
forcing the disk back into the orbital plane (Petterson 1977).

In principle, for every system a combination of viscosity and
accretion efficiency can be found which makes the system stable,
unstable or highly unstable against radiation driven warping. A deeper
knowledge of the internal physics of accretion disk is therefore
required to determine the cause of the warping. Recently, Ogilvie \&
Dubus (2001) found that radiation driven warping in LMXB becomes
relevant when the orbital period is above $\sim$1\,day, which makes it
possibly important for \grs.

Whether or not a warped accretion disk is the relevant cause of the
long term periodicity in \grs\ remains uncertain, because the
590\,days period is visible in the non-thermal part of the X-ray
spectrum and in the radio flux, but not in the disk
parameters. However, it is worth mentioning two issues: (i) a possibly
existing long term variability in the soft X-ray flux and the disk
parameters can easily be masked by the high short term variability in
timing and spectral behavior of \grs. (ii) The observed periodicity
can originate due to variability of the mass accretion rate or the
viscosity of the inner part of the disk. One can assume that the Roche
lobe overflow of the donor star deposits it's mass at different disk
radii, depending on the mass transfer rate, thus differing mass
density. The radial flow in the inner part of the accretion disk is
therefore not constant, leading to a periodic variation of the amount
of soft seed photons being comptonized in the corona. Therefore, the
electron temperature in the corona varies on the same time scale due
to Compton cooling, as the power law slope. The amount of matter to be
ejected in a jet changes with similar periodicity which explains the
overall periodicity in the radio flux.

In the same way, periodic instabilities in the secondary affect the
amount of accreted matter and may therefore cause or at least
influence the observed periodicity. Wind-driven limit cycles (Shields
\etal\ 1986) as discussed for LMC X-3 (Wilms \etal\ 2001) can also
exist in \grs. In order to drive a wind from the outer parts of the
accretion disk the sound speed for the inverse Compton temperature,
$kT_{IC}$, has to exceed the escape velocity in the outer disk. With
an orbital separation of $a\sim7.5\cdot 10^{12}$\,cm the outer disk
radii is likely of the order of $\sim5\cdot 10^{12}$\,cm. Thus, \grs\
fullfils the criterium given in Equ.(3) of Wilms \etal\ (2001) for
$kT_{IC}>1$\,keV and may therefore be driving a substantial
Compton-heated wind.

A warped accretion disk would temporarily shadow the irradiated donor
star. A search for the 590\,days periodicity in the IR flux of the
donor would also help in the understanding of the origin and mechanism
of the long term behavior of \grs.

\section{Conclusion}

We present the discovery of a 590\,days long term periodicity in the
power law slope of the hard X-ray spectrum, the hard X-ray flux and
the radio flux (at several frequencies) of the microquasar \grs. No
such periodicity is seen in the soft X-ray component. We discussed the
observed behavior in the context of a warped accretion disk and of
wind-driven limit cycles. We have found, that a periodic mass density
variation in the disk, produced by the warp and/or by periodic
instabilities in the donor, can describe the observations, by
producing a varying amount of soft seed photons for the Comptonization
process and for the relativistic mass ejecta, very well. A wind-driven
limit cycles is also a possible origin of the 590\,days periodicity.

\begin{acknowledgements}
The authors thank E. H. Morgan (MIT) and G. G. Pooley (MRAO) for
providing the ASM and {\it RT} data and W. A. Heindl ({\it RXTE}) and
K. A. Arnaud (XSPEC) for their comments. We thank R. Schwarz (AIP) for
the kind help with the analysis of variance. {\it The Ryle Telescope}
is supported by PPARC. The {\it Green Bank Interferometer} is a
facility of the National Science Foundation operated by the NRAO in
support of NASA High Energy Astrophysics programs.
\end{acknowledgements}

\end{document}